# Electrostatic properties and current transport of two-dimensional Schottky barrier diode

**Fangbo Xu, Alex Kutana, Yang Yang, Boris I. Yakobson (✉)**

Department of Materials Science & NanoEngineering, Rice University, Houston, TX77005

(✉) biy@rice.edu

## Abstract

Recently demonstrated metal-semiconductor heterojunctions with few-atom thickness show their promise as 2D Schottky contacts for future integrated circuits and nanoelectronics. The theory for 3D Schottky contacts, however, fails on these low-dimensional systems. Here, we propose a new model that yields carrier distribution and potential profile across the 2D metal-semiconductor heterojunction under the equilibrium condition, based on the input from first-principle calculations. Our calculation suggests that, at the same forward bias, the current density of a stack of 2D graphene-phosphorene Schottky diodes may be ten thousand times higher than that of a traditional 3D Schottky diode and offer less energy dissipation.

Schottky barrier diodes (SBD) have been indispensable in power-rectification and radio-frequency applications[1,2,3,4,5,6,7]. As three-dimensional integrated circuits, which vertically stack layers of electronic components, has been widely recognized in recent years[8,9,10], 2D materials hold great promise[11,12,13]. In particular, some new 2D semiconductors, such as phosphorene and monolayer transition-metal dichalcogenides ($MoS_2$, $WSe_2$, etc.), have moderate band gaps (1-2 eV) and high carrier mobilities[14,15,16], and may be used to form 2D Schottky contacts. Seamless lateral 2D heterojunctions of graphene and 2D semiconductors have been demonstrated experimentally by employing conventional photolithography[17,18].

However, the laws governing 3D SBDs fail qualitatively in low-dimensional cases. For instance, it erroneously predicts non-trivial electric field outside the space charge region (SCR)[19]. Also, first-principle calculations proved to be prohibitively expensive to study energy band offsets [20,21,22]. To this end, we derive a generic model of a 2D SBD and evaluate its electrostatics and I-V curves (The system-defined carrier-capturing edge states are neglected). Then we apply this model to characterize the novel features of a 2D SBD formed by graphene and phosphorene.

In the 3D SBD model, the electric field becomes trivial outside a well-defined SCR. When it becomes few-atom thick, however, the surface charges give rise to a field that vanishes only at infinity. Imagine a heterojunction of two semi-infinite metal and semiconductor sheets lying at $z=0$, if an electron is moved from $x=-\infty$ (metal) to $x=+\infty$ (semiconductor) along an electric field line, the work in this process makes up the difference of Fermi levels. Since the path is an immense semi-circle normal to the plane, we have $E_z(x) \sim 1/x$ and hence the surface charge density $\sigma_s(x) \sim 1/x$, resulting in an indefinite total charge[23]. Accordingly, in this work, metal and semiconductor are assumed to be strips with widths of 1 μm and 2 μm, respectively.

Previous work demonstrates that charged graphene behaves like classical conductor with a constant potential profile[24]. By conformal transformation[25], the charge density $\sigma_m(x)$ of metal strip induced by a coplanar semiconductor strip with $\sigma_s(x)$ is evaluated as:

$$\sigma_m(x') = -\int_{x_1}^{x_2} \frac{\sigma_s(x)}{\pi} \frac{\sqrt{x^2 - a^2}}{(x - x')\sqrt{a^2 - x'^2}} dx \qquad (1)$$

assuming the metal strip lies at $z=0$ and $x \in [-a, a]$, while the semiconductor strip lies between $[x_1, x_2]$, ($x_1 > a$). Suppose $\sigma_s(x) > 0$, due to electric neutrality, in metal there is another positive charge with the equal amount, and its distribution, evaluated by $\sim (a^2 - x^2)^{-1/2}$, ensures equipotential across the metal strip[3]. Also, the semiconductor injects electrons which obey the same distribution, neutralizing the positive charge. Therefore, Eq(1) gives the net charge on the metal strip.

Figure 1(a) shows the band diagram before and after metal and *n*-type semiconductor reach equilibrium. The work function of semiconductor is assumed lower than metal. At equilibrium, we have:

$$\begin{aligned} qV_{b0} &= \Delta E_{F\_m} + \Delta E_{F\_s} \\ &= \Delta E_{F\_m} + q\Phi + q\phi_n^* - q\phi_n \end{aligned} \qquad (2)$$

in which $\Phi$ is the electrostatic potential relative to the interface. For a given total transferred charge, $Q_{sc}$, the semiconductor itself may achieve equilibrium as long as the carrier density satisfies

$$n(x) = n(x_1) \exp[q\Phi(x)/kT] \qquad (3)$$

where $\Phi(x_1) = 0$. With an initial guess of $\Phi(x)$, one may obtain $\sigma_s(x)$, and then $\sigma_m(x)$ by Eq(1), and further a new $\Phi(x)$ as well as $\Delta E_{F\_m}$. When $\Phi(x)$ and $n(x)$ converge, we use Eq(2) to check if the junction achieves equilibrium, if negative, $Q_{sc}$ needs to be adjusted until it equilibrates the heterojunction.

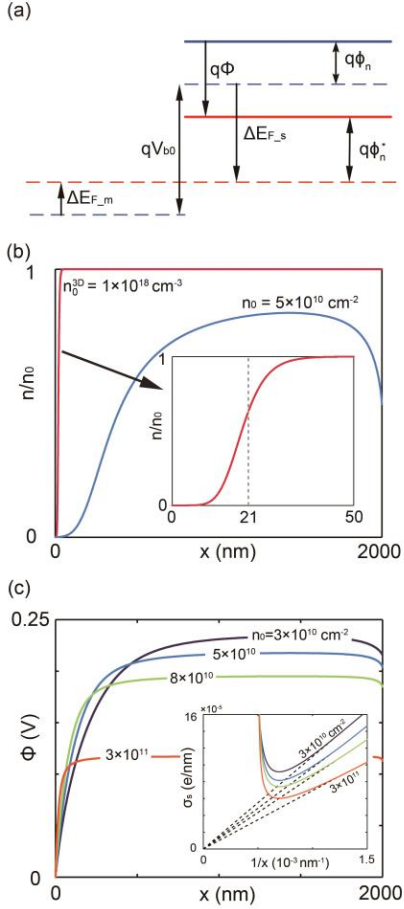

**Figure 1. (a) Band diagram of a heterojunction of metal and *n*-type semiconductor. (b) Carrier density across the semiconductor in 3D and 2D junctions. The thickness of the silicon strip is assumed to be 0.5 nm. Inset: zoom-in of 3D semiconductor near the interface, and the dashed line marks the boundary of the space charge region. (c) The electrostatic potential profile across the semiconductor strip in a 2D W-Si junction. Inset: charge distribution near the far end of the semiconductor strip.**

To highlight the impact of merely lowering dimensionality, we work with an imaginary 0.5 nm thick SBD of tungsten and *n*-type silicon, while the bulk properties (e.g. Schottky barrier, permittivity) are retained. Figure 1(b) shows the change of $n(x)$ in silicon at $n_0^{3D} = 1 \times 10^{18}$ cm$^{-3}$ in contrast to 3D model, and the inset displays the zoom-in near the 3D interface. We find that in the 2D SBD the neutral region outside SCR no longer exists and the whole semiconductor becomes charged. Figure 1(c) suggests that $\Phi(x)$ exhibits a zero-field plateau. Here we take the height of the plateau as the built-in potential barrier $V_{bi}$. In the 3D model, we have $V_{bi} = \Phi_B - \phi_n$, which barely varies since $\phi_n$ is almost invariant if $n_0$ is high. In our 2D model, however, $V_{bi}$ significantly decreases as $n_0$ increases. Deep insight reveals that, at a higher $n_0$, more electrons pour into the metal, causing a larger $\Delta E_{F\_m}$ and hence a lower $V_{bi}$. In addition, the inset of Figure 1(c) displays that, at the distance far away from the interface (excluding the edge effect), the charge distribution behaves approximately $\sim 1/x$, which is consistent with our argument on the infinite 2D SBD.

Next we apply our model to a 2D SBD of graphene and phosphorene. Doping of phosphorene has been explored both theoretically and experimentally recently[26, 27, 28, 29], but the doping mechanism is not yet well established. Therefore we employ the conventional nearly-free electron model involving 2D anisotropy, see Supplementary Information (S.I.). Necessary parameters regarding electrostatics are listed in Table 1.

**Table 1 Parameters for electrostatic properties of phosphorene. $\varepsilon_z$ = 5.27. The effective masses and carrier mobilities are extracted from Ref. 12. The permittivities in each direction, independent of carrier type, are fitted by the model of a parallel plate capacitor by varying the thickness of vacuum slab.**

| | $m^*/m_0$ | $\mu$ ($10^3$ cm$^2$V$^{-1}$s$^{-1}$) | $\varepsilon/\varepsilon_0$ | $\Phi_B$ (eV) | MFP (nm) |
|---|---|---|---|---|---|
| **armchair (e)** | 0.17 | 1.10-1.14 | 15.23 | ~1.50 | ~30 |
| **zigzag (e)** | 1.12 | ~0.08 | 11.62 | 1.0 | ~6 |
| **armchair (h)** | 0.15 | 0.64-0.70 | - | ~ 0.1 | ~17 |
| **zigzag (h)** | 6.35 | 10-26 | - | 0.5 | ~3300 |

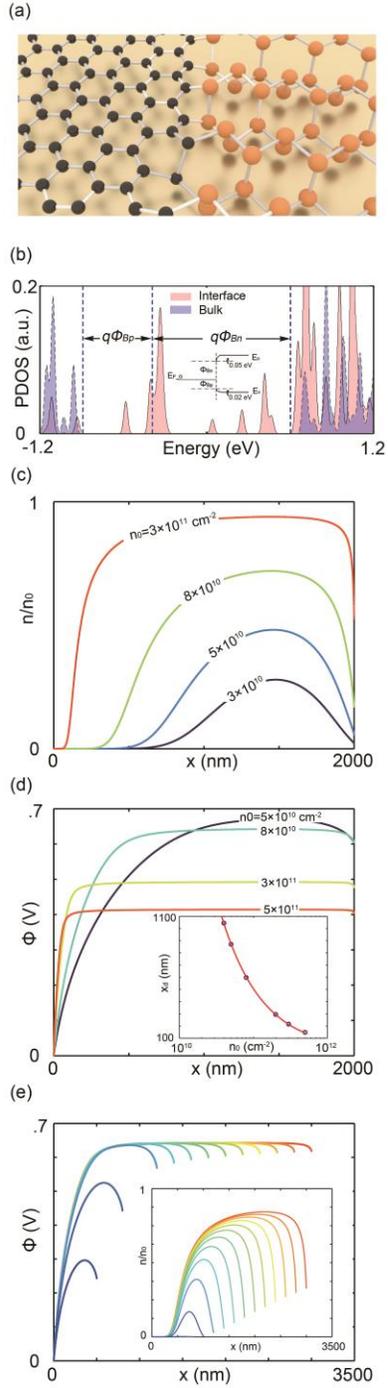

**Figure 2** The electrostatic properties of the 2D graphene-phosphorene heterojunction, with widths of 1 μm and 2 μm, respectively. The transport is along the zigzag direction of phosphorene. (a) The atomic structure. (b) The density of states projected on individual P atoms located at the interface (pink) and far from the interface (blue). $E_{F\_G}$ marks the position of the Fermi level of the graphene. (c) The carrier density across the semiconductor strip at various doping concentrations. (d) Potential profile at various doping concentrations. Inset shows dependence of $x_d$ on the doping level. (e)

**Potential profile and carrier density (inset) at a variety of semiconductor widths at $n_0 = 8 \times 10^{10}$ cm$^{-2}$, $x_d$=1.07 μm.**

The graphene-phosphorene (G-P) heterojunction is modeled as periodic arrays of dislocations[30]. Figure 2(a) shows a possible structure of C(5,1)|P(3,3) with 0.8% lattice mismatch, accommodating the boundary perpendicular to the zigzag direction of phosphorene. Meanwhile, C(2,1)|P(2,0) is also created with 1.1% lattice mismatch for the boundary perpendicular to the armchair direction (see Figure 1 of S.I.).

The Schottky barrier for *n*- and *p*-type SBD, namely $\Phi_{Bn}$ and $\Phi_{Bp}$, are determined in the following way. We perform density functional theory (DFT) calculation on graphene alone to obtain its Fermi level relative to the lower bound of the projected density of states (PDOS), and then locate it on the PDOS of P atoms at the interface, as illustrated by Figure 2(b). Note that the DFT calculations were performed on intrinsic semiconductor under 0 K, so the carrier transfer is negligible, and accordingly the band bending is ascribed to the interfacial dipoles. In this paper we ignore the impact on $\Phi_B$ due to carrier transfer, since the local bending of bands due to the carrier transfer is estimated less than 30 meV ($< 1\% \, \Phi_B$) for all doping levels.

Figure 2(c) displays the carrier density in the *n*-doped phosphorene with a variety of doping concentrations. Compared with the 2D W-Si junction, due to the higher $\Phi_B$, $n(x)$ has a less steep transition, a narrowed plateau and more significant edge effect. Similar features are also found in the potential profile, as presented in Figure 2(d). We define the distance from the interface where $E_x < 10^{-5}$ V/nm as the width of SCR, $x_d$, and have $V_{bi} = \Phi(x_1 + x_d) - \Phi(x_1)$. Referring to the 3D model in which $x_d^{3D} \sim \left(n_0^{3D}\right)^{-.5}$, Figure 2(d) inset gives $x_d^{2D} \sim \left(n_0^{2D}\right)^{-.77}$. This faster decay agrees with the fact that increasing $n_0$ lowers $V_{bi}$ as explained in the W-Si example.

The semiconductor strip needs to be wider than SCR since otherwise $V_{bi}$ becomes unpredictable. Figure 2(e) shows the results for various widths of the phosphorene strip, $w_s$. Apparently $\Phi(x)$ becomes flat after it reaches $V_{bi}$ if $w_s > x_d$, while otherwise the potential barrier always exhibits lower than $V_{bi}$. One may also find that $V_{bi}$ stays invariant irrespective of $w_s$, and this indicates that $Q_{sc}$, which determines $\Delta E_{F\_m}$, also becomes invariant. Thus the diverse $w_s$ gives distinct carrier distributions, as shown in the inset of Figure 2(e).

Furthermore, the mechanism of equilibration turns out different as the number of carriers varies. With a given $n_0$, altering $w_s$ essentially modifies the number of carriers, which determines upper limit of $V_{bi}$. When the amount of carriers is so adequate that the carrier transfer hardly changes $\phi_n$, the potential $\Phi(x)$ accounts for the majority of $\Delta E_{F\_s}$. However, if carriers are too few, the increase of $\phi_n$ becomes dominant in $\Delta E_{F\_s}$, and the semiconductor is almost drained under this circumstance.

**Currents**

As indicated in Table 1, in phosphorene the mean free path (MPF) of electrons is so trivial that the classical model for current transport is still applicable. The formalism of diffusion current[31] is readily modified to adapt to our 2D model and gives:

$$J_{n\_dif} = \mu_n kTn(x_2) \frac{1-\exp\left(-\frac{qV_a}{kT}\right)}{\int_{SCR} \exp\left[-\frac{q\Phi^*(x)}{kT}\right]dx} \quad (4)$$

where $\Phi^*(x)$ is the potential profile in the presence of the bias $V_a$ applied on the ends of the semiconductor strip (see S.I.).

To develop the expression for the thermionic current of 2D SBD, we utilize the anisotropic nearly-free electron model and eventually reach (see S.I. for derivation):

$$J_{n\_thm} = A^* T^{3/2} \exp\left(-\frac{q\Phi_{Bn}}{kT}\right)\left[\exp\left(\frac{qV_a}{kT}\right)-1\right] \quad (5)$$

where $A^* = \frac{qk\sqrt{2\pi km_y^*}}{2\pi^2\hbar^2}$ is the 2D Richardson constant. Note that the 3D thermionic current holds the similar form but varies as $\sim T^2$. Our calculation shows that the diffusion current is much higher than the thermionic current in G-P SBD (see S.I.)

On the other hand, the holes possess an appreciable MPF (~3 μm) in the zigzag direction. We estimate the tunneling current with the non-equilibrium Green's function approach (NEGF) combined with DFT calculations[32], along with the WKB approximation. The NEGF+DFT approach may give the transmission coefficient, $M(E)$, which characterizes the intrinsic tunneling defined by the interfacial dipoles and coupling of Bloch waves. The WKB approximation is used to estimate the hole tunneling probability through the barrier across the SCR as $\Theta(E) = \exp\left[-\int_{SCR} \frac{2\sqrt{2m_x^*(q\Phi^* - E)}}{\hbar}dx\right]$. Thus the total transmission coefficient is evaluated as $T(E) = M(E)\Theta(E)$, and further the tunneling current is obtained by Landauer formula[33]. Figure 3 shows the tunneling current of holes in the zigzag direction, and $M_{s \to m}(E)$, from the intrinsic phosphorene to the graphene (inset). Under an increasing forward bias but lower than $V_{bi}$, the current is found to increase exponentially. Higher doping concentration yields a shrunk $x_d$ and a lowered $V_{bi}$, which jointly produce a raised $\Theta$.

Compared with the *n*-type SBD, the *p*-type SBD yields higher current under the same condition, implying a much lower turn-on voltage. For instance, at $n_0 = 5 \times 10^{10}$ cm$^{-2}$ and $V_a = 0.1$ V, $J_{p\_tnl} = 6.2$ µA/cm, higher than $J_{n\_dif}$ by 6 magnitude orders. If the 2D Schottky diodes are stacked up with a spacing of 2 nm, at 0.1 V the current density could be $10^4$ times higher than that of a traditional 3D W-Si junction (2.9 mA/cm$^2$)[34]. Additionally, the *p*-type SBD employs ballistic transport of carriers, promising low energy dissipation. Note that the tunneling from metal to semiconductor is neglected, due to the absence of states in the bandgap of semiconductor.

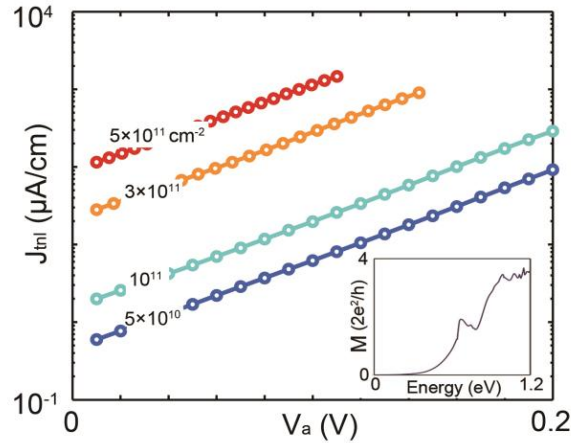

Figure 3 The quantum transport of holes in the zigzag direction of phosphorene under forward bias. Inset shows the intrinsic transmission coefficient of the graphene-phosphorene junction in the absense of carriers at zero-bias regime.

In summary, we present a generic model for 2D SBD which shows that the charge distribution and potential profile are subject to pronounced modification relative to the classical 3D model. Moreover, the graphene-phosphorene heterojunction is studied in terms of electrostatics and current transport arising from diverse mechanisms. Specially, due to the considerable MFP of holes, the ballistic current density is dramatically higher than 3D model, and therefrom a promising quantum 2D SBD with low turn-on voltage and less energy dissipation is revealed.

Methods

First-principle calculations have been carried out with density functional theory implemented in VASP. In particular, the Perdew–Burke–Ernzerhof (PBE) exchange-correlation functional is used to relax the atomistic structure with the maximum force less than 0.01 eV/Å on each atom and the vacuum slab larger than 15 Å, while HSE06 hybrid functional is employed for electronic structures, predicting a band gap of 1.58 eV for phosphorene. The intrinsic transmission coefficients through the junction of graphene and zigzag-oriented phosphorene were computed using non-equilibrium Green's function formalism implemented in the TRANSIESTA code[35], averaged on multiple **k**-points perpendicular to the transmission direction in the zero-bias regime, including calculation of self-energies describing the coupling of the scattering region with a semi-infinite graphene lead and a semi-infinite phosphorene lead.

Acknowledgement